\documentclass[12pt]{article}
\usepackage[english]{babel}
\usepackage[dvips]{epsfig}
\usepackage{cite}
\newcommand{\Kmax}{9} 
\newcommand{\ep}{\varepsilon}
\newcommand{\Li}[2]{{\mbox{Li}}_{#1\!}\left(#2\right)}
\newcommand{\Cl}[2]{{\mbox{Cl}}_{#1}\left(#2\right)}

\newcommand{\Ls}[2]{{\mbox{Ls}}_{#1}\!\left(#2\right)}
\newcommand{\LS}[3]{{\mbox{Ls}}_{#1}^{(#2)}\!\left(#3\right)}

\newcommand{\tfrac}[2]{{\textstyle{\frac{#1}{#2}}}}
\newcommand{\Snp}[2]{{\it{S}}_{#1\!}\left(#2\right)}
\newcommand{\ST}[2]{\left[ #1 \atop #2\right]} 

\begin{document}

\renewcommand{\thefootnote}{\fnsymbol{footnote}}
\thispagestyle{empty}
\begin{flushright}
\end{flushright}
 \vspace*{2.0cm}
 \begin{center}
 {\large \bf
 {\sf lsjk} --  a C++ library for arbitrary-precision numeric evaluation of the generalized log-sine functions }
 \end{center}
 \vspace*{2.0cm}
\begin{center}
M.Yu.~Kalmykov~\footnote{Supported by RFBR grant \# 04-02-17192 
 and the Heisenberg-Landau Programme. \\
 Email address: kalmykov@thsun1.jinr.ru}
\quad and \quad 
A.~Sheplyakov~\footnote{Supported by RFBR grant \#~02-02-16889, 
by grant  of the ministry of science and technology policy of
the Russian Federation \#~2339.2003.2, 
DFG grant 436~RUS~113/626/0-1 and the Heisenberg-Landau Programme \\
 E-mail: varg@thsun1.jinr.ru}
\\

\vspace*{2.0cm}

Bogoliubov Laboratory of Theoretical Physics, \\
Joint Institute for Nuclear Research, \\
$141980$ Dubna (Moscow Region), Russia

\end{center}

\begin{abstract}
Generalized log-sine functions $\LS{j}{k}{\theta}$ appear in
higher order $\ep$-expansion of different Feynman diagrams. We
present an algorithm for the numerical evaluation of these functions
for real arguments. This algorithm is implemented as a C++ library with
arbitrary-precision arithmetics for integer $ 0 \leq k \leq \Kmax$
and $j \geq 2$. Some new relations and representations of the 
generalized log-sine functions are given.
\end{abstract}

\newpage 

{\bf\large PROGRAM SUMMARY}
\vspace{4mm}

\begin{sloppypar}
\noindent   {\em Title of program\/}: {\sf lsjk} \\[2mm]
   {\em Version\/}: 1.0.0
   {\em Release\/}: 1.0.0
   {\em Catalogue number\/}: \\[2mm]
   {\em Program obtained from\/}:
   {\tt http://thsun1.jinr.ru/\~{}varg/dist/} \\[2mm]
   {\em E-mail: varg@thsun1.jinr.ru} \\[2mm]
   {\em Licensing terms\/}: GNU General Public Licence \\[2mm]
   {\em Computers\/}: all \\[2mm]
   {\em Operating systems\/}: POSIX \\[2mm]
   {\em Programming language\/}: {\tt C++    } \\[2mm]
   {\em Memory required to execute\/}: Depending on the complexity of
	 the problem, at least 32Mb RAM recommended.\\[2mm]
   {\em Other programs called\/}: The CLN library for
     arbitrary-precision arithmetics is required
     at version 1.1.5 or greater. \\[2mm] 
   {\em External files needed\/}: none \\[2mm]
   {\em Keywords\/}:  Generalized log-sine functions, Feynman integrals, Polylogarithms. \\[2mm]
   {\em Nature of the physical problem\/}: 
        Numerical evaluation of the generalized log-sine functions
        for real argument in the region $0 < \theta < \pi $.
				These functions appear in Feynman integrals. \\[2mm]
   {\em Method of solution\/}: 
        Series representation for the real argument in the region $0 < \theta < \pi$. \\[2mm]
   {\em Restriction on the complexity of the problem}: Limited up to $\LS{j}{\Kmax}{\theta}$, 
        and $j$ is an arbitrary integer number.
        Thus, all function up to the weight 12 in the region $0 < \theta < \pi$ can be
	evaluated. The algorithm can be extended up to higher values of 
        $k$ $(k > \Kmax)$ without modification. \\[2mm]
   {\em Typical running time}:  Depending on the complexity of problem.
	 See Table~\ref{BenchT} in section \ref{BenchNCheck}\\[2mm]
\end{sloppypar} 

\newpage

{\bf\large LONG WRITE-UP}
\vspace{4mm}

\renewcommand{\thefootnote}{\arabic{footnote}}
\setcounter{footnote}{0}

\section{Introduction}
\label{Sintro}
\setcounter{equation}{0}
Quite recently it has been realized that generalized log-sine
functions \cite{Lewin} play an important role in analytical
calculations of multiloop Feynman diagrams. Within dimensional
regularization \cite{dimreg} with an arbitrary space-time dimension 
$d=4-2\ep$, the log-sine functions $\Ls{j}{\theta}$
appear in the construction of the {\em all} order
$\ep$-expansion of the one-loop propagator type diagrams with
arbitrary masses \cite{Crete}, three-point integrals with massless
internal lines and arbitrary (off-shell) external momenta
\cite{D-ep}, two-loop vacuum diagrams with arbitrary masses
\cite{D-ep,DT2} and two-loop single scale diagrams of propagator type
\cite{FKK99}.  The appearance of the generalized log-sine functions
$\LS{j}{k}{\theta}$ $(k=1,2)$ was first detected in the
finite part of the three-loop bubble master-integrals \cite{FK99,DK1}. 
More examples of the one-,
two- and three-loop Feynman diagrams whose $\ep$-expansion contains
the generalized log-sine functions can be
found in \cite{DK1,ls3_pi/2,JKV2003,DK_tokyo,top,DK2}.  The
generalized log-sine functions are also related to the derivatives
of the generalized hypergeometric functions with respect to their
parameters \cite{DK2,MK2004} and to the multiple Euler--Zagier 
sums \cite{FK99,DK1}. A collection of the known analytical 
properties of these functions can be found in
Appendix~A of \cite{DK1} (see also \cite{DK_bastei}).

Here we present C++ code for arbitrary precision numeric 
evaluation of the generalized function $\LS{j}{k}{\theta}$
of the real argument $0 < \theta < \pi$ and $k \leq \Kmax$.
For calculation of the $\Ls{j}{\theta}$ and
$\LS{j}{1}{\theta}$ functions there exists a FORTRAN program 
(see the description in \cite{KV00}).

The plan of the paper is the following. 
Section~\ref{Sdefnprops} contains the definition and some 
properties of the generalized log-sine functions.
Section~\ref{SstupidAlgo} is the description of 
the {\sf lsjk} library. 
In Appendix \ref{appendixTRIGONOMETIRC} we present 
relations between  the functions $\LS{j}{0,1,2}{\theta}$ and 
infinite series containing the trigonometric functions and  
harmonic sums.

\section{Generalized log-sine functions}
\label{Sdefnprops}
\setcounter{equation}{0}
\subsection{Definition and reflection symmetries}

The generalized log-sine functions are defined as \cite{Lewin}
\begin{equation}
\LS{j}{k}{\theta} =   - \int\limits_0^\theta {\rm d}\phi \;
   \phi^k \ln^{j-k-1} \left| 2\sin\frac{\phi}{2}\right| \, , \quad 
\Ls{j}{\theta} = \LS{j}{0}{\theta} \; , 
\label{log-sine-gen}
\end{equation}   
where $k,j$ are integer numbers, $k \geq 0$ and $j \geq k+1$ and
$\theta$ is an arbitrary real number.  

According to the definition (\ref{log-sine-gen}), the following
integral and differential relations hold:
\begin{eqnarray}
&& 
\frac{1}{\alpha^k} \LS{j\!+\!k}{i\!+\!k}{\alpha \theta} \!=\! \theta^k \LS{j}{i}{\alpha \theta}
\!-\! k \int\limits_0^\theta {\rm d}\phi\;
{\phi}^{k-1} \LS{j}{i}{\alpha \phi} 
\; , 
\\ && 
\frac{{\rm d}}{{\rm d} \theta} \LS{j\!+\!k}{i\!+\!k}{\alpha \theta} 
\!=\! ( \alpha \theta) ^k \frac{{\rm d}}{{\rm d} \theta} \LS{j}{i}{\alpha \theta} \; ,
\label{integral_ls}
\end{eqnarray}
where $\alpha$ is a real number.  In particular, 
$$
\LS{j}{j-1}{\theta} = -\frac{1}{j} \theta^j \;. 
$$
It is easy to obtain a relation between functions of the opposite arguments
\begin{equation}
\LS{j}{k}{-\theta}  = (-1)^{k+1} \LS{j}{k}{\theta} \;, \quad 
\theta > 0 \; . 
\label{sym}
\end{equation}
Another relation was deduced in Ref.~\cite{Lewin}
via the integral representation: 
\begin{equation}
\LS{n}{r}{2 m \pi \!-\! \theta} 
= 
\LS{n}{r}{2 m \pi } 
\!+\! (-1)^{r-1} \LS{n}{r}{\theta} 
\!-\! \sum_{p=1}^r (-1)^{r-p} \left( 2 m \pi \right)^p
{r \choose p} \LS{n-p}{r-p}{\theta } \;,
\label{rel1}
\end{equation}
where $m$ is an integer number.  Using the symmetry property
(\ref{sym}),
this relation can be written as
\begin{equation}
\LS{n}{r}{2 m \pi \!+\! \theta} 
=   
\LS{n}{r}{2 m \pi } 
\!+\!\LS{n}{r}{\theta} 
\!+\! \sum_{p=1}^r (-1)^{r-p} \left( 2 m \pi \right)^p
{r \choose p} \LS{n-p}{r-p}{\theta } \;.
\label{rel2}
\end{equation}
\noindent
The generalized log-sine functions appear in the
decomposition of polylogarithms $\Li{j}{1\!-\!e^{{\rm i}\theta}}$
into real and imaginary parts (see Eqs.~(A.7) in \cite{DK1}).  
It is easy to show that the generalized Nielsen polylogarithms
\cite{Nielsen} $S_{a,b}(z)$ of a unit-circle complex argument $z=e^{{\rm i}
\theta}$ also reduce to a combination of the generalized log-sine
integrals (\ref{log-sine-gen})~\footnote{The corresponding equation in
Ref.~\cite{DK_tokyo} contains an misprint in the last term.  We are
grateful to A.~Kotikov for correspondence.}
\begin{eqnarray}
&& \hspace{-5mm}
\Snp{a,b}{e^{{\rm i} \theta}} = \frac{{\rm i }^{a} (-1)^b}{(a-1)! \,b!}
\int_0^\theta
{\mbox d} \phi \; 
( \theta \!-\! \phi )^{a-1}
\left[ \ln \left| 2\sin\tfrac{\phi}{2} \right|
\!-\! \tfrac{1}{2}{\rm i} (\pi \!-\! \phi)
\right]^b 
\!+\! \sum_{k=0}^{a-1} \frac{({\rm i} \theta)^k}{k!} \Snp{a-k,b}{1} 
\nonumber \\ && \hspace{-5mm}
= \frac{({\rm -i })^{a} (-1)^b}{(a-1)! \,b!}
\sum_{k=0}^{a-1} \sum_{j=0}^b \sum_{m=0}^j 
(-\theta)^{a-1-k} (-\pi)^{j-m}
\left(\frac{{\rm i}}{2} \right)^j
\left( a-1 \atop k \right) 
\left( b   \atop j \right) 
\left( j   \atop m \right) 
\LS{m+k+b+1-j}{m+k}{\theta}
\nonumber \\ && \hspace{-5mm}
+ \sum_{k=0}^{a-1} \frac{({\rm i} \theta)^k}{k!} \Snp{a-k,b}{1} 
\; .
\end{eqnarray}
\subsection{Identities between the generalized log-sine and Clausen functions}
It was shown in \cite{DK1} that the 
function $\LS{k+2}{k}{\theta}$ is always expressible
in terms of the Clausen functions $\Cl{j}{\theta}$.
This follows from the fact that $\Ls{2}{\theta}=\Cl{2}{\theta}$,
the differential identity (\ref{integral_ls}), which 
in this case has the following form:
\begin{equation}
\frac{{\rm d}}{{\rm d} \theta} \LS{k\!+\!2}{k}{\theta} 
\!=\! \theta^k \frac{{\rm d}}{{\rm d} \theta} \Cl{2}{\theta} \; ,
\label{part1}
\end{equation}
and the integration rules for the Clausen function \cite{Lewin},
\[
\Cl{2n}{\theta} =  \int\limits_0^\theta {\rm d}\phi\; \Cl{2n-1}{\phi} ,
\quad
\Cl{2n+1}{\theta} =  \zeta_{2n+1}
- \int\limits_0^\theta {\rm d}\phi\; \Cl{2n}{\phi}.
\]
The general solution of Eq.~(\ref{part1}) could be represented in the 
following form 
\begin{eqnarray}
\LS{2j\!+\!2}{2j}{\theta} & =  & 
\sum_{k=0}^{2j}
\frac{(2j)!}{(2j-k)!} \theta^{2j-k} \Cl{2+k}{\theta} (-1)^{k(k-1)/2} \; , 
\\ 
\LS{2j\!+\!3}{2j+1}{\theta} & =  & 
(-1)^j (2j+1)! \left[ \Cl{2j+3}{\theta}  \!-\! \zeta_{2j+3} \right]
\nonumber \\ && \hspace{2mm}
+ \sum_{k=0}^{2j}
\frac{(2j+1)!}{(2j \!+\! 1 \!-\! k)!} \theta^{2j \!+\! 1 \!-\! k} \Cl{2+k}{\theta} (-1)^{k(k-1)/2} \; .
\end{eqnarray}
The corresponding expressions up to $\LS{6}{4}{\theta}$ have been presented in \cite{DK1}.
\subsection{Identities between the generalized log-sine functions and the generalized Nielsen polylogarithms}
In Ref.~\cite{DK1} the following relations between 
the log-sine functions and generalized Nielsen polylogarithms were deduced: 
\begin{eqnarray}
\label{Ls<->S}
&& \hspace*{-7mm}
{\rm i} \sigma \left[ \Ls{j}{\pi}  - \Ls{j}{\theta} \right]
= \frac{1}{2^j j}  \ln^j (-z) \left[ 1 - (-1)^j \right] 
\nonumber \\ && 
+ (-1)^j (j-1)! 
\sum_{p=0}^{j-2} \frac{\ln^p (-z)}{2^p p!} 
\sum_{k=1}^{j-1-p} (-2)^{-k}
\left[ S_{k,j-k-p}(z) - (-1)^p S_{k,j-k-p}(1/z) \right],
\end{eqnarray}
where the relation between $z$ and $\theta$ is defined via
\begin{equation}
\label{def_z} 
z \equiv e^{ {\rm i} \sigma \theta}, \hspace{5mm}
\ln(-z-{\rm i}\sigma 0) = \ln(z) - {\rm i} \sigma \pi , \hspace{5mm}
\sigma=\pm 1 , 
\end{equation}
and $0 \leq \theta \leq \pi $.  For higher values of $j$ the number of
generalized polylogarithms involved in Eq.~(\ref{Ls<->S}) can be
reduced (see Eqs.~(2.21), (2.22) in \cite{DK1}).  
The relation between $\LS{j\!+\!1}{1}{\theta}$
and the generalized Nielsen polylogarithms is also known
(see Eq.~(A.20) in \cite{DK1}), 
\begin{eqnarray}
&&  
\LS{j+1}{1}{\theta} - \LS{j+1}{1}{\pi}
= \theta \Biggl [ \Ls{j}{\theta} - \Ls{j}{\pi}
\Biggr ]
- \frac{1}{2^j j (j+1)}  \ln^{j+1} (-z) \left[ 1 - (-1)^j \right]
\nonumber \\ && 
- (-1)^j (j-1)! \sum_{p=0}^{j-2} \frac{\ln^p (-z) }{2^p p!}
\sum_{k=1}^{j-1-p} \frac{(-1)^k}{2^k} k 
\left[ S_{k+1,j-k-p}(z) + (-1)^p S_{k+1,j-k-p}(1/z) \right ] 
\nonumber \\ &&
+ 2 (-1)^j (j-1)! \sum_{k=1}^{j-1} \frac{(-1)^k}{2^k} k  S_{k+1,j-k}(-1) \; .
\label{LS{j}{1}}
\end{eqnarray}
Using the algorithm described in \cite{DK1} we found the following relation 
for $\LS{j\!+\!2}{2}{\theta}$ and generalized Nielsen polylogarithms:
\begin{eqnarray}
&&  
{\rm i}  \sigma \left[ \LS{j+2}{2}{\pi}  - \LS{j+2}{2}{\theta} \right]   
= 
2 {\rm i}  \sigma \theta \Biggl[ \LS{j+1}{1}{\pi} - \LS{j+1}{1}{\theta} \Biggr]
- {\rm i}  \sigma \theta^2 \Biggl [ \Ls{j}{\pi} - \Ls{j}{\theta} \Biggr ]
\nonumber \\ &&
- \frac{1}{2^{j-1} j (j+1)(j+2)}  \ln^{j+2} (-z) \left[ 1 - (-1)^j \right]
 \nonumber \\ &&
- (-1)^j (j-1)! \sum_{p=0}^{j-2} \frac{\ln^p (-z) }{2^p p!}
\sum_{k=1}^{j-1-p} \frac{(-1)^k k(k \!+\! 1)}{2^k}  
\left[ S_{k+2,j-k-p}(z) - (-1)^p S_{k+2,j-k-p}(1/z) \right ] 
\nonumber \\ &&
+ 4 (-1)^j (j-1)!  \ln (-z) \sum_{k=1}^{j-1} \frac{(-1)^k}{2^k} k  S_{k+1,j-k}(-1) \; .
\label{LS{j}{2}}
\end{eqnarray}
The procedure, described in \cite{DK1}, can be repeated several times, so that 
the generalized log-sine functions with arbitrary indices can be directly 
related to Nielsen polylogarithms.

\subsection{Series representation}
In the region $0 < \theta < \pi$, the 
series representation of the generalized log-sine functions
directly follows from the definition~(\ref{log-sine-gen}) 
by  substituting $y=\sin\tfrac{\phi}{2}$
\begin{equation}
\LS{j}{k}{\theta} = - 2^{k+1} \int\limits_0^{\sin(\theta/2)} 
\frac{{\rm d}y}{\sqrt{1-y^2}}
\left(\arcsin y \right)^k \ln^{j-k-1}(2y). 
\end{equation}
Using integration by parts, this integral can be rewritten in the following form:
\begin{eqnarray}
\hspace{-5mm}
\LS{j}{k}{\theta} =  
- \frac{\theta^{k+1}}{k\!+\!1} \ln^{j-k-1}\left(2 \sin \frac{\theta}{2} \right)
\!+\! \frac{2^{k+1}}{k\!+\!1} (j\!-\!k\!-\!1) \int\limits_0^{\sin(\theta/2)} 
\frac{{\rm d}y}{y} \left(\arcsin y \right)^{k+1} \ln^{j-k-2}(2y). 
\label{integralII}
\end{eqnarray}
Expanding $\left(\arcsin y \right)^k$ in $y$ (see below) 
and using 
\[
\int x^a \ln^b x ~{\rm d} x = (-1)^b\; b!\; x^{a+1} 
\sum_{p=0}^b \frac{\left(- \ln x \right)^{b-p}}{(b-p)!(a+1)^{p+1}} \; ,
\]
we obtain the multiple series representation. 
The generating function  for the coefficients of the Taylor expansion of 
$\left(\arcsin y \right)^k$ and their explicit form 
for $k=1,2,3,4$ can be found in \cite{RN1}.
We have calculated the coefficients of the Taylor expansion for $k=5,\cdots,12$. 
For even powers, the Taylor expansion of $\left(\arcsin y \right)^{2k}$ 
can be represented in the following form: 
\begin{eqnarray}
\frac{\left(\arcsin y \right)^{2k}}{(2k)!}
= \sum_{m=1}^\infty \frac{[(m-1)!]^2}{(2m)!} 4^{m-1} y^{2m} B(m,k) \; , 
\label{arcsin:even}
\end{eqnarray}
where the coefficients $B(m,k)$ are expressible in terms of harmonic sums
\begin{equation}
\label{harmonic}
S_a(n) = \sum_{k=1}^n \frac{1}{k^a} \;. 
\end{equation}
For the lowest values of $k$ we have
\begin{eqnarray}
&& 
B(m,1) = 1 \; , 
\quad 
B(m,2) = \sigma_1 \; , 
\quad 
B(m,3) = \frac{1}{2}  ~\left[\sigma_1^2 - \sigma_2 \right] \; , 
\nonumber \\ && 
B(m,4) =  \frac{1}{3!} ~\left[\sigma^3_1 - 3 \sigma_1  \sigma_2  +  2 \sigma_3 \right] \; , 
\nonumber \\ && 
B(m,5) =  \frac{1}{4!} ~\left[
\sigma_1^4 
+ 8 \sigma_1 \sigma_3
+ 3 \sigma_2^2 
- 6 \sigma_4
- 6 \sigma_1^2 \sigma_2
\right] \; , 
\nonumber \\ && 
B(m,6) = \frac{1}{5!} ~\left[ 
\sigma_1^5
\!-\! 30 \sigma_1 \sigma_4 
\!-\! 20 \sigma_2 \sigma_3 
\!+\! 20 \sigma_1^2 \sigma_3 
\!-\! 10 \sigma_1^3 \sigma_2 
\!+\! 15 \sigma_1 \sigma_2^2  
\!+\! 24 \sigma_5 
\right] \; ,
\label{b_nk}
\end{eqnarray}
and 
\begin{eqnarray}
\sigma_a = \sum_{j=1}^{m-1} \frac{1}{(2j)^{2a}} \equiv \frac{1}{4^a} S_{2a}(m-1) \; . 
\end{eqnarray}
For odd powers, the Taylor expansion of $\left(\arcsin y \right)^{2k+1}$ 
has the following structure: 
\begin{eqnarray}
\frac{\left(\arcsin y \right)^{2k+1}}{(2k+1)!}
= \sum_{m=0}^\infty \left( 2m \atop m\right) \frac{y^{2m+1}}{4^m(2m+1)} C(m,k+1) \; , 
\label{arcsin:odd}
\end{eqnarray}
where the coefficients $C(m,k)$ also expressible in terms of harmonic sums, 
\begin{eqnarray}
&& 
C(m,1) = 1 \; , 
\quad 
C(m,2) = h_1 \; , 
\quad 
C(m,3) = \frac{1}{2}  ~\left[h_1^2 - h_2 \right] \; , 
\nonumber \\ && 
C(m,4) =  \frac{1}{3!} ~\left[h^3_1 - 3 h_1  h_2  +  2 h_3 \right] \; , 
\nonumber \\ && 
C(m,5) =  \frac{1}{4!} ~\left[
h_1^4 
+ 8 h_1 h_3
+ 3 h_2^2 
- 6 h_4
- 6 h_1^2 h_2
\right] \; , 
\nonumber \\ && 
C(m,6) = \frac{1}{5!} ~\left[ 
h_1^5
\!-\! 30 h_1 h_4 
\!-\! 20 h_2 h_3 
\!+\! 20 h_1^2 h_3 
\!-\! 10 h_1^3 h_2 
\!+\! 15 h_1 h_2^2  
\!+\! 24 h_5 
\right] \; ,
\label{c_nk}
\end{eqnarray}
and
\begin{eqnarray}
h_a = \sum_{j=1}^m \frac{1}{(2j-1)^{2a}} \equiv S_{2a}(2m-1) - \frac{1}{4^a} S_{2a}(m-1) \; . 
\end{eqnarray}
Starting with Eq.~(\ref{integralII}) and using expansions 
(\ref{arcsin:even}) and (\ref{arcsin:odd}), we get 
\begin{eqnarray}
&& \hspace{-7mm}
\LS{j}{2k}{\theta} = 
(-1)^j \left( 2 \sin \tfrac{\theta}{2} \right)
2^{2k} (2k)! (j-2k-1)! \sum_{p=0}^{j\!-\!2k\!-\!1} 
\frac{\left[ - \ln \left( 2 \sin \tfrac{\theta}{2} \right) \right]^p}{p!}
\nonumber \\ && \hspace{10mm}
\times 
\sum_{m=0}^\infty \left( 2m \atop m\right) 
\frac{ \left( 2 \sin \tfrac{\theta}{2} \right)^{2m}}{16^m (2m+1)^{j-2k-p}} C(m,k+1) \; , 
\label{LS:even}
\\ && \hspace{-7mm}
\LS{j}{2k+1}{\theta} = 
(-1)^{j-1} 2^{2k+2} (2k+1)! (j\!-\!2k\!-\!2)! \sum_{p=0}^{j-2k-2} 
\frac{\left[ - \ln \left( 2 \sin \tfrac{\theta}{2} \right) \right]^p}{p!}
\nonumber \\ && \hspace{10mm}
\times 
\sum_{m=1}^\infty \frac{1}{\left( 2m \atop m\right) }
\frac{ \left( 2 \sin \tfrac{\theta}{2} \right)^{2m}}{(2m)^{j-2k-p}} B(m,k+1) \; , 
\label{LS:odd}
\end{eqnarray}
where $k=0,1,2,3,4,5$ and the coefficients $B(m,k)$ and $C(m,k)$ are given 
in (\ref{b_nk}) and (\ref{c_nk}), respectively.

The special interest in the calculation of the multiloop Feynman diagrams 
\cite{B99,FK99,DK1,DK_tokyo}
is related to the value of the generalized log-sine functions at 
the ``sixth root of unity'',  $\theta = \frac{\pi}{3}$. 
In this case $\ln \left( 2 \sin \tfrac{\theta}{2} \right) = 0$ and only 
the $p=0$ term in the finite sum in Eqs.~(\ref{LS:even}),(\ref{LS:odd}) survives, 
\begin{eqnarray} 
&& \hspace{-7mm}
\LS{j}{2k}{\tfrac{\pi}{3}} = 
(-1)^j 2^{2k} (2k)! (j\!-\!2k\!-\!1)! 
\sum_{m=0}^\infty  \left( 2m \atop m\right) 
\frac{C(m,k+1)}{16^m (2m+1)^{j-2k}} \;, 
\\ && \hspace{-7mm}
\LS{j}{2k+1}{\tfrac{\pi}{3}} = 
(-1)^{j-1} 2^{2k+2} (2k+1)! (j\!-\!2k\!-\!2)! 
\sum_{m=1}^\infty \frac{1}{\left( 2m \atop m\right) }
\frac{B(m,k+1)}{(2m)^{j-2k}} \; . 
\end{eqnarray}
Similar series for $\Ls{j}{\theta}$ and $\LS{j}{1}{\theta}$ 
have been presented earlier in \cite{KV00}.

\subsection{Special values of the argument}
\label{special_values}

The values of the generalized log-sine functions of the argument 
$\theta=\pi$ and $2 \pi$ can be related with the values of 
some well-known special functions. The relations of this type 
were discussed in Lewin's book \cite{Lewin}. 
For completeness we present some of them below.
The values of $\Ls{j}{\pi}$ can be expressed in terms of 
the Rieman $\zeta$-function, for any $j$
(see Eqs.~(7.108), (7.110) in \cite{Lewin}), 

\begin{eqnarray}
&& \hspace{-15mm}
\Ls{n+1}{\pi} =  
- \pi \left( \frac{{\rm d}}{{\rm d}x} \right)^n 
\left.\frac{\Gamma(1 \!+\! x)}{\Gamma^2(1 \!+\! \tfrac{x}{2})} \right|_{x=0}
= 
- \pi \left( \frac{{\rm d}}{{\rm d} x} \right)^n \exp
\Biggl[\sum_{m=2}^\infty \frac{x^m}{m} (-1)^m \left(1 \!-\! 2^{1-m} \right) \zeta_m
\Biggr]_{x=0} 
\end{eqnarray}
so that 
\[
\Ls{2}{\pi} = 0, \quad\!\!\!
\Ls{3}{\pi} =  -\tfrac{1}{2} \pi \zeta_2, \quad\!\!\!
\Ls{4}{\pi} =  \tfrac{3}{2} \pi \zeta_3, \quad\!\!\!
\Ls{5}{\pi} = -\tfrac{57}{8} \pi \zeta_4, \quad\!\!
\Ls{6}{\pi} = \tfrac{45}{2} \pi \zeta_5
+ \tfrac{15}{2} \pi \zeta_2 \zeta_3 ,
\]
etc. 
Putting in Eq.~(\ref{rel1}) $r=0$ and $\theta = 2 \pi (m-1)$, 
solving the difference equation and taking into account
the relation between values of function $\Ls{j}{\theta}$ at 
$\theta = \pi$ and $2\pi$, we finally get
\begin{eqnarray}
\Ls{n}{2 m \pi}  & = & 2 m \Ls{n}{\pi} \; .
\label{ls:2pi}
\end{eqnarray}
From relation (\ref{rel1}) for $r=1$ and $\theta = 2 \pi m$ 
we have a new relation 
\begin{eqnarray}
\LS{n}{1}{2 \pi m}  & = & 2 m^2 \pi \Ls{n-1}{\pi} \; .
\label{LS:1:2pi}
\end{eqnarray}
Putting $r=3$ and $\theta = 2 \pi m$ in Eq.~(\ref{rel1}) 
and using Eqs.~(\ref{ls:2pi}) (\ref{LS:1:2pi}) we get
\begin{eqnarray}
\LS{n}{3}{2 \pi m}  & = & 3 \pi m \LS{n-1}{2}{2 \pi m} - 4 m^4 \pi^3 \Ls{n-3}{\pi} \; ,
\label{LS:3:2pi}
\end{eqnarray}
where the value of $\LS{m}{2}{2 \pi m}$ should be calculated
independently.  The integral $\LS{n}{m}{2 \pi}$ can be expressed
in terms of a combination of $\zeta$-functions (see Eq.~(7.136)
in \cite{Lewin}):
\begin{eqnarray}
\hspace{-5mm}
\LS{n+m+1}{n}{2\pi} & = & 
\!-\! 
2 \pi  (- {\rm i})^n
\left( \frac{{\rm d}}{{\rm d} x} \right)^n 
\left( \frac{{\rm d}}{{\rm d} y} \right)^m 
\left. e^{{\rm i} x \pi}
\frac{\Gamma(1 \!+\! y)}
     {\Gamma(1 \!+\! \tfrac{y}{2} \!+\! x)
      \Gamma(1 \!+\! \tfrac{y}{2} \!-\! x)} \right|_{x\!=\!y\!=\!0} \; . 
\end{eqnarray}
It can be written in the following form~\footnote{The term 
$-[1 \!+\! (-1)^{n}] (n-1)! \zeta_n$ is absent in Eq.~(7.139)
of \cite{Lewin}.}:
\begin{eqnarray}
\hspace{-7mm} 
\LS{n+m+1}{n}{2\pi} & = &   
\!-\! 
2 \pi (- {\rm i})^n
\left( \frac{{\rm d}}{{\rm d} x} \right)^n 
\left( \frac{{\rm d}}{{\rm d} y} \right)^m 
\nonumber \\ && \hspace{-25mm}
\times 
\exp \Biggl[ {\rm i} x \pi
+ \sum_{p=2}^\infty \frac{y^p}{p!} (-1)^p \left(1 \!-\! 2^{1-p} \right) \zeta_p
\nonumber \\ && \hspace{-5mm}
- \sum_{q=2}^\infty \frac{x^q}{q!} [1 \!+\! (-1)^{q}] 
\Biggl\{ (q-1)! \zeta_q \!+\! 
\sum_{p=1}^\infty \frac{y^p}{p!} \left(\frac{-1}{2}\right)^p \Gamma(p+q) \zeta_{p+q} 
\Biggr\}
\Biggr]_{x=y=0} \; ,
\end{eqnarray}
so that for $n=2$ we get~\footnote{In Eq.~(7.141) of \cite{Lewin},
the term  $2 \pi^2 \Ls{r+1}{\pi}$ should read $\frac{8}{3} \pi^2 \Ls{r+1}{\pi}$.}
\begin{eqnarray}
\hspace{-5mm}
\LS{m+3}{2}{2\pi} & = & \frac{8}{3} \pi^2 \Ls{m+1}{\pi}
\!-\! \pi (m\!+\!1)! (-2)^{2-m} \zeta_{2+m}
\nonumber \\ && 
+ 4 m! \sum_{p=1}^{m-1} \frac{(-2)^{p-m}}{p!} (m\!-\!p\!+\!1)\Ls{p+1}{\pi} \zeta_{2+m-p} \;,
\quad m \geq 2 \; .
\label{LS_2_2pi}
\end{eqnarray}
In particular, we have~\footnote{The correct result for 
$\LS{5}{2}{2 \pi}$ was first given in \cite{DK1}.}  
\begin{equation}
\LS{5}{2}{2 \pi} = - \frac{13}{45} \pi^5 \; , \quad 
\LS{6}{2}{2 \pi} = 
5 \pi^3 \zeta_3 \!+\! 12 \pi \zeta_5 \; , \quad 
\LS{7}{2}{2 \pi} =
- \frac{29}{105} \pi^7 \!-\! 24 \pi \zeta_3^2 \; . 
\label{LS_2_2pi_particular}
\end{equation}
Substituting expression (\ref{LS_2_2pi}) in 
Eq.~(\ref{LS:3:2pi}) we get $(m=1)$
\begin{eqnarray}
\hspace{-5mm}
\LS{n}{3}{2\pi} & = &  4 \pi^3 \Ls{n-3}{\pi}
- 3 \pi^2 (n-3)! (-2)^{6-n} \zeta_{n-2}
\nonumber \\ && 
+ 3 \pi (n-4)! \sum_{p=1}^{n-5} \frac{(-2)^{p-n+6}}{p!} (n\!-\!p\!-\!3) \Ls{p+1}{\pi} \zeta_{n-2-p} \;,
\quad n \geq 5 \; , 
\label{LS_3_2pi}
\end{eqnarray}
and 
\begin{equation}
\LS{5}{3}{2 \pi} = 12 \pi^2 \zeta_3 \; , \quad 
\LS{6}{3}{2 \pi} = 
- \frac{8}{15} \pi^6  \; , \quad 
\LS{7}{3}{2 \pi} =
9 \pi^4 \zeta_3 + 36 \pi^2 \zeta_5 \; . 
\label{LS_3_2pi_particular}
\end{equation}

The values of the generalized log-sine functions of the argument $\pi$
are related to the values of the $\Gamma$-functions and generalized
Nielsen polylogarithms of the argument $z=-1$ (see section 7.9.9
in \cite{Lewin}).  For lower values of $k~(k=0,1,2)$ the
proper relations can be deduced from Eqs.~(\ref{Ls<->S}) --
(\ref{LS{j}{2}}):
\begin{eqnarray}
&& \hspace{-5mm}
\Ls{j}{\pi}
= 
\frac{{\rm i}^{j+1}}{2^j j}  \pi^j  \left[ 1 \!-\! (-1)^j \right]
\nonumber \\ && 
+ (-1)^j (j-1)! \sum_{p=0}^{j-2} \frac{{\rm i}^{p+1}}{p!}\left(\frac{\pi}{2} \right)^p
\sum_{k=1}^{j-1-p} \frac{(-1)^k}{2^k} S_{k,j-k-p}(1) \left[ 1 - (-1)^p \right ] 
\; , 
\label{LS{j}{0}{pi}}
\\ && \hspace{-5mm}
\LS{j+1}{1}{\pi}
= 
\frac{{\rm i}^{j+1}}{2^j j (j+1)}  \pi^{j+1}  \left[ 1 \!-\! (-1)^j \right]
- 2 (-1)^j (j-1)! \sum_{k=1}^{j-1} \frac{(-1)^k}{2^k} k  S_{k+1,j-k}(-1) \; .
\nonumber \\ && 
+ (-1)^j (j-1)! \sum_{p=0}^{j-2} \frac{ \left( {\rm i} \pi \right)^p }{2^p p!}
\sum_{k=1}^{j-1-p} \frac{(-1)^k}{2^k} k  S_{k+1,j-k-p}(1) \left[ 1 + (-1)^p \right ] 
\; , 
\label{LS{j}{1}{pi}}
\\ &&  \hspace{-5mm}
\LS{j+2}{2}{\pi} 
= 
-\frac{{\rm i}^{j+3}}{2^{j-1} j (j+1)(j+2)}  \pi^{j+2} \left[ 1 \!-\! (-1)^j \right]
\!-\! 4 \pi (-1)^j (j \!-\! 1)!  \sum_{k=1}^{j-1} \frac{(-1)^k}{2^k} k  S_{k+1,j-k}(-1) 
\nonumber \\ &&
- (-1)^j (j-1)! \sum_{p=0}^{j-2} \frac{{\rm i}^{p+1}}{p!} \left(\frac{\pi}{2} \right)^p
\sum_{k=1}^{j-1-p} \frac{(-1)^k k(k \!+\! 1)}{2^k}  S_{k+2,j-k-p}(1) 
\left[ 1 - (-1)^p \right ] \; .
\label{LS{j}{2}{pi}}
\end{eqnarray}
In particular, we have 
\begin{eqnarray}
\LS{4}{1}{\pi} & = & -2 \Snp{2,2}{-1} \!-\! \tfrac{11}{8} \zeta_4 \; , 
\\ 
\LS{5}{1}{\pi} & = & 6 \Snp{2,3}{-1} \!+\! \tfrac{3}{2} \zeta_2 \zeta_3 \!-\! \tfrac{105}{32} \zeta_5 \; , 
\quad 
\LS{5}{2}{\pi}  =  2 \pi \LS{4}{1}{\pi} \!+\! 2 \pi \zeta_4 \;, 
\end{eqnarray}
where the values of generalized Nielsen polylogarithms of argument $z = \pm 1$
can be extracted from \cite{Nielsen,Nielsen2}. For completeness we present the proper results below
\begin{eqnarray}
\Snp{a,b}{1} & = & \Snp{b,a}{1} \; , \quad 
\Snp{2,2}{1} =  \tfrac{1}{4} \zeta_4 \;, \quad 
\Snp{1,3}{1} = \zeta_4 \; , \quad 
\Snp{2,3}{1} = 2 \zeta_5 - \zeta_2 \zeta_3 \; ,
\nonumber \\ 
\Snp{1,3}{-1} & = & 
\tfrac{1}{24} \ln^4 2 
- \tfrac{1}{4} \zeta_2 \ln^2 2 
+ \tfrac{7}{8} \zeta_3 \ln 2
- \zeta_4 
+ \Li{4}{\tfrac{1}{2}} \; , 
\nonumber \\ 
\Snp{2,3}{-1} & = & 
- \tfrac{1}{15} \ln^5 2 
\!+\! \tfrac{1}{3} \zeta_2 \ln^3 2 
\!-\! \tfrac{7}{8} \zeta_3 \ln^2 2 
\!+\! \tfrac{1}{2} \zeta_2 \zeta_3 
\!+\! \tfrac{33}{32} \zeta_5
\!-\! 2 \Li{4}{\tfrac{1}{2}} \ln 2 
\!-\! 2 \Li{5}{\tfrac{1}{2}} \; . 
\nonumber \\ 
\Snp{2,2}{-1} & = &  
2 \Snp{1,3}{-1}
+  \tfrac{1}{8} \zeta_4 \; , \quad 
\Snp{3,2}{-1} = \tfrac{1}{2} \zeta_3 \zeta_2 - \tfrac{29}{32} \zeta_5 \; .
\label{Nielsen_1}
\end{eqnarray}
\section{The {\sf lsjk} library}
\label{SstupidAlgo}
\setcounter{equation}{0}

\subsection{Domain}
Our algorithm for the numerical evaluation of the generalized log-sine
functions in the region $0 < \theta < \pi$ is based on the
multiple series representation (\ref{LS:even}) and (\ref{LS:odd}).
The explicit value of the generalized log-sine function at $\theta =
\pi$ and $\theta = 2 \pi$ can be extracted by the approach described
in Lewin's book \cite{Lewin} (see also section \ref{special_values}).
For an argument belonging to the region  $\pi < \theta < 2\pi$ the
following representation is valid
\begin{equation}
\LS{j}{k}{\theta} = (-1)^{k+1}
\sum_{p=0}^k (-2\pi)^p {k \choose p} 
\Biggl\{ 
\LS{j-p}{k-p}{2\pi-\theta} \!-\! \LS{j-p}{k-p}{2\pi}
\Biggr\} \; . 
\end{equation}
For negative values of argument, $\theta < 0$,  the
reflection symmetry (\ref{sym})  is applied. 
Eqs.~(\ref{rel1}) and (\ref{rel2}) allow us to express the value of 
the generalized log-sine function of the argument $|\theta| > 2 \pi$ in terms 
of the function $\LS{k}{j}{\theta}$ at $0 < \theta < \pi$,  $\pi$ and $2 \pi$.

In the present version of {\sf lsjk} we have considered only 
the region $0 < \theta < \pi$.

\subsection{Description}
For arbitrary-precision arithmetics, {\sf lsjk} uses the CLN library
\footnote{CLN can be downloaded from http://www.ginac.de/CLN.
Alternatively, you may wish to check your favored operating system
distribution for a precompiled package.}. CLN has a rich set of
number classes:
\begin{itemize}
\item Integers (with unlimited precision)
\item Rational numbers (with unlimited precision)
\item Floating-point numbers:
        \begin{itemize}
                \item Short float
                \item Single float
                \item Double float
                \item Long float numbers (with unlimited precision)
        \end{itemize}
\item Complex numbers
\end{itemize}
and implements many elementary functions on these numbers.

{\sf lsjk} provides two functions \footnote{cl\_R is CLN type for real numbers,
see CLN manual for details.}
\begin{verbatim}
const cl_R Ls(const unsigned & j, const cl_R &x);
const cl_R Ls(const unsigned & j, const unsigned & k, const cl_R &x);
\end{verbatim}
These functions represent $\Ls{j}{x}$ and $\LS{j}{k}{x}$,
respectively.  The argument $x$ is supposed to satisfy $0 < x <\pi$,
parameter $k$ should be $0 \leq k \leq \Kmax$~\footnote{In the
present version we have implemented only first five coefficients
$B(m,k)$ and $C(m,k)$, $k=1,2,3,4,5$ from Eqs.~(\ref{b_nk}) and
(\ref{c_nk}), respectively.}.  {\sf lsjk} can be used as any regular
library, e.g. by writing programs and linking executables to it.
\subsection{Installation instructions}
\label{UNIXforDummies}
The installation procedure is reduced to the simple three 
steps\footnote{last step may require superuser privileges} 
of
\begin{verbatim}
./configure && make && make install 
\end{verbatim}
The ``configure'' script can be given a number of options to enable
and disable various features. A few of the more important ones are
documented in the ``INSTALL'' file from the {\sf lsjk} distribution.

After installation {\sf lsjk} can be used as any regular library,
e.g. by writing programs and linking executables with it.
{\sf lsjk} includes shell script ``lsjk-config'' that can be
used for setting compiler and linker command-line options required to
compile and link a program with the {\sf lsjk} library. Thus, a simple 
program can be compiled as
\begin{verbatim}
c++ -o simple `lsjk-config --cppflags` simple.cpp `lsjk-config --libs`
\end{verbatim}
See ``README'' file from the {\sf lsjk} distribution for more details.
\subsection{Benchmarks and cross-checks}
\label{BenchNCheck}
Correctness of implementation was cross-checked by comparison
with results of numerical integration \footnote{For numerical
integration we used GNU scientific library, see 
http://www.gnu.org/software/gsl} of Eq.~(\ref{log-sine-gen})
for $4 \leq j \leq 11$, $0 \leq k \leq \Kmax$, with  
randomly chosen argument $0 < x < \pi$.  
We also checked that the results of the evaluation of the
functions $\Ls{j}{\theta}$ and $\LS{j}{1}{\theta}$ coincide
with the corresponding results of the FORTRAN program
described in \cite{KV00}.

To give a reader a rough idea about the evaluation time,
Table \ref{BenchT} may be helpful. It should be noted that
the calculation time for values of the argument near $\pi$
increases substantially.

\begin{table}
\begin{tabular}{|c|c|c|c|c|c|c|c|c|c|c|c|}
\hline
{} & \multicolumn{3}{|c|}{$j=4$} & \multicolumn{4}{|c|}{$j=5$} & \multicolumn{4}{|c|}{$j=6$} \\
\hline
$d \setminus k$ & $0$ & $1$ & $2$ & $0$ & $1$ & $2$ & $3$ & $0$ & $1$ & $2$ & $3$  \\
\hline
128      &0.053  &0.034  &0.055  &0.055  &0.037  &0.058  &0.048  &0.059  &0.04   &0.063  &0.052  \\
\hline
256      &0.166  &0.106  &0.187  &0.172  &0.111  &0.195  &0.16   &0.177  &0.118  &0.203  &0.166  \\
\hline
512      &0.685  &0.448  &0.872  &0.699  &0.47   &0.876  &0.742  &0.727  &0.491  &0.9    &0.741  \\ 
\hline
1024     &3.212  &2.219  &5.084  &3.294  &2.292  &4.459  &4.013  &3.327  &2.342  &4.496  &3.786  \\
\hline
\end{tabular}
\caption{Dependence of the average calculation time (in second), as reported by {\rm getrusage}(2), 
of $\LS{j}{k}{2\pi/3}$ on the precision (d) of evaluation.
It was measured on a Duron/800MHz with 128Mb RAM.}
\label{BenchT}
\end{table}

\subsection{Example of a program}
The following program evaluates $\LS{5}{2}{2\pi/3}$ up to 1024 digits.
\begin{verbatim}
#include <iostream>
#include <cln/cln.h>
#include <lsjk/lsjk.hpp>
using namespace std;
using namespace cln;

int main(int argc, char** argv)
{
        const unsigned j=5;
        const unsigned k=2;
        float_format_t prec=float_format(1024); // precision
        cl_R ThePi = pi(prec);
        cl_R res = Ls(j, k, 2*ThePi/3);
        cout << res << endl;
        return 0;
}
\end{verbatim}
The output of this program is:
\begin{verbatim}
-0.5181087868296801173472656387316967550218796682431532140673894724824\
6493059206791506817591796234263409228316887407062572713789701522832828\
8301238053344434601555482416349687142642605456956152340876808788330125\
2744524532005650653916633546607642565939433325023687049969640726184300\
7710801944912063838971724384311449565205834807350617442006639919209369\
6654189591396805453280242324416887003772837420727727140290932114280662\
5550331483934346157017999680014851656153847980079446225103428769237020\
8915162715787613074342208607956842722678511943161304156600252481686319\
3071905942805038395349632095408925090936865076348021184023670239583644\
8060572488328604823250125773056264305964195471500442032760480010663468\
6808425894080311707295955974893312124168605505481096916643118747707399\
7726995685152764342004787909957595709564369273416476440874928297302997\
3226232121625055566818301147295999943567467361944097333638320437023447\
2148571069312485138137784268429835976279026626925221417350916538992701\
34033573580034488504210827748452303627506494601119041L0
\end{verbatim}

\section{Conclusion}
\label{Soutro}

In this paper we have described the C++ library {\sf lsjk}
for arbitrary-precision numerical evaluation of the generalized
log-sine functions for the real argument $0 < \theta < \pi$.
The evaluation is based on the series representations (\ref{LS:even})
and  (\ref{LS:odd}).  The present version allows one to evaluate
generalized log-sine functions up to $\LS{j}{\Kmax}{\theta}$, thus,
all functions up to weight 11 can be evaluated. The algorithm can be
extended to higher weights without modification. In particular,
implementation of the next coefficients $B(m,6)$ and $C(m,6)$
(see Eqs.~(\ref{b_nk}) and (\ref{c_nk})) allows one to evaluate
all functions up to weight 14.

Explicit formulae (\ref{LS{j}{2}}) relating $\LS{j}{2}{\theta}$
and the generalized Nielsen polylogarithms have been
obtained.  Using the relations of this type for the functions
$\Ls{j}{\theta}$ and $\LS{j}{1}{\theta}$, 
the relations (\ref{LS{j}{0}:series})--(\ref{LS{j}{2}:series}) 
between series included trigonometric functions and  
harmonic sums and the functions $\LS{j}{0,1,2}{\theta}$ 
have been deduced.  
As a particular case, some
new identities (\ref{zoo_begin})-(\ref{zoo:2}) for a special type of harmonic sums
have been obtained.  Another application of these expressions is
the relation between the values of functions $\LS{j}{1,2}{\theta}$
at $\theta = \pi$ and the Nielsen polylogarithms at $z= \pm 1$
(\ref{LS{j}{1}{pi}}), (\ref{LS{j}{2}{pi}}).  The correct expressions
for the functions $\LS{j}{2,3}{\theta}$ at $\theta = 2 \pi$ in
terms of the $\Gamma$-functions and their derivatives are obtained
(\ref{LS_2_2pi}), (\ref{LS_3_2pi}).

\noindent
{\bf Acknowledgements.}
We are grateful to A.~Davydychev and S.~Mikhailov for useful discussions
and A.~Davydychev and G.~Sandukovskaya for careful reading of the manuscript. 
Many thanks to Editor for his suggestions and help.
%
%
%
\appendix

\section{Identities between $\LS{j}{0,1,2}{\theta}$ and  infinite series}
\label{appendixTRIGONOMETIRC}
\setcounter{equation}{0}
Identities between the generalized log-sine functions
and infinite series involving
trigonometric functions and harmonic sums
follow  from the power series expansion of the generalized 
Nielsen polylogarithms
\begin{eqnarray}
\Snp{n,p}{z} = \sum_{k=1}^\infty  \ST{k}{p}\frac{z^k}{k! k^n} \; ,  
\label{Nielsen}
\end{eqnarray}
where $\ST{k}{p}$ denotes Stirling numbers of first kind defined 
as \cite{Stirling}
$$
\ln^p (1-z) =  (-1)^p p! \sum_{k=p}^\infty  \ST{k}{p}\frac{z^k}{k!} \;.
$$
Using  Eqs.~(\ref{Ls<->S}) and (\ref{Nielsen}) we get 
\begin{eqnarray}
&& \hspace*{-7mm}
\Ls{j}{\theta}  - \Ls{j}{\pi} 
= 
\frac{1}{2^j j}  (-1)^{j-1} {\rm i}^{j-1} (\pi-\theta)^j \left[ 1 - (-1)^j \right] 
\nonumber \\ && 
- 
(-1)^j (j-1)! 
\sum_{p=0}^{j-2} \frac{(-{\rm i} \sigma)^{p+1}(\pi-\theta)^p}{2^p p!} 
\sum_{k=1}^{j-1-p} (-2)^{-k}
\sum_{n=1}^\infty \ST{n}{j\!-\!k\!-\!p}\frac{1}{n! n^k} 
\nonumber \\ && 
\times 
\left\{  
\cos (n \theta) \left[ 1 \!-\! (-1)^p \right] 
\!+\! {\rm i} \sigma \sin (n \theta) \left[ 1 \!+\! (-1)^p \right] 
\right\}.
\label{LS{j}{0}:series}
\end{eqnarray}
Similar expansion can be deduced for $\LS{j}{1}{\theta}$ and $\LS{j}{2}{\theta}$ with the help 
of Eqs.~(\ref{LS{j}{1}}) and (\ref{LS{j}{2}}),
\begin{eqnarray}
&&  
\LS{j+1}{1}{\theta} - \LS{j+1}{1}{\pi}
= \theta \Biggl [ \Ls{j}{\theta} - \Ls{j}{\pi}
\Biggr ]
- \frac{ {\rm i}^{j+1}}{2^j j (j+1)} (\pi-\theta)^{j+1} \left[ 1 - (-1)^j \right]
\nonumber \\ && 
- (-1)^j (j-1)! \sum_{p=0}^{j-2} \frac{(-{\rm i} \sigma)^p(\pi-\theta)^p}{2^p p!}
\sum_{k=1}^{j-1-p} \frac{(-1)^k}{2^k} k 
\sum_{n=1}^\infty \ST{n}{j\!-\!k\!-\!p}\frac{1}{n! n^{k+1} }
\nonumber \\ && 
\times 
\left\{  
\cos (n \theta) \left[ 1 \!+\! (-1)^p \right] 
\!+\! {\rm i} \sigma \sin (n \theta) \left[ 1 \!-\! (-1)^p \right] 
\right\}
\nonumber \\ &&
+ 2 (-1)^j (j-1)! \sum_{k=1}^{j-1} \frac{(-1)^k}{2^k} k  S_{k+1,j-k}(-1) \; ,
\label{LS{j}{1}:series}
\\ &&  
\LS{j+2}{2}{\theta} - \LS{j+2}{2}{\pi}
= 
2 \theta \Biggl [ \LS{j+1}{1}{\theta} - \LS{j+1}{1}{\pi} \Biggr ]
- \theta^2 \Biggl [ \Ls{j}{\theta} - \Ls{j}{\pi} \Biggr ]
\nonumber \\ && 
+ \frac{{\rm i}^{j+3}}{2^{j-1} j (j+1)(j+2)} (\pi-\theta)^{j+2} \left[ 1 - (-1)^j \right]
\nonumber \\ && 
+ (-1)^j (j-1)! \sum_{p=0}^{j-2} \frac{(-{\rm i} \sigma)^{p+1} (\pi-\theta)^p}{2^p p!}
\sum_{k=1}^{j-1-p} \frac{(-1)^k}{2^k} k(k+1) 
\sum_{n=1}^\infty \ST{n}{j\!-\!k\!-\!p}\frac{1}{n! n^{k+2} }
\nonumber \\ && 
\times 
\left\{  
\cos (n \theta) \left[ 1 \!-\! (-1)^p \right] 
\!+\! {\rm i} \sigma \sin (n \theta) \left[ 1 \!+\! (-1)^p \right] 
\right\}
\nonumber \\ &&
+ 4  (\pi-\theta) (-1)^j (j-1)! \sum_{k=1}^{j-1} \frac{(-1)^k}{2^k} k  S_{k+1,j-k}(-1) \; ,
\label{LS{j}{2}:series}
\end{eqnarray}
For lower values of $j$ these expressions have the following form:
\begin{eqnarray}
&&  \hspace*{-7mm}
\Ls{3}{\theta} = -2 \sum_{n=1}^\infty \frac{\sin (n \theta)}{n^2} S_1
- \tfrac{1}{12} \theta \left( \theta^2 - 3 \pi \theta + 3 \pi^2 \right) \; ,
\label{Ls3_series}
\\ && \hspace*{-7mm}
\Ls{4}{\theta}
= 3 \sum_{n=1}^\infty \frac{\sin (n \theta)}{n^2} 
\left[ S_1^2 \!-\! S_2 
\!+\! \tfrac{1}{4} (\pi \!-\! \theta)^2 
\!-\! \tfrac{1}{2} \frac{1}{n^2} 
\right]
\!-\! \tfrac{3}{2} (\pi \!-\! \theta) \sum_{n=1}^\infty \frac{\cos (n \theta)}{n^3} 
\!+\! \tfrac{3}{2} \pi \zeta_3 \; , 
\label{Ls4_series}
\\ && \hspace*{-7mm}
\Ls{5}{\theta} = 
3 (\pi-\theta) \sum_{n=1}^\infty \frac{\cos (n \theta)}{n^2} \left[ S_1^2 \!-\! S_2 \right]
+ 3 \sum_{n=1}^\infty \frac{\sin (n \theta)}{n^3} \left[ S_1^2 \!-\! S_2 \right]
\nonumber \\ && \hspace*{-2mm}
- 4 \sum_{n=1}^\infty \frac{\sin (n \theta)}{n^2} \left[ S_1^3 \!-\! 3 S_1 S_2 \!+\! 2 S_3 \right]
\!-\! \tfrac{1}{16} \pi \theta (\pi - \theta) (\pi^2 \!-\! \pi \theta \!+\! \theta^2)
\!-\! \tfrac{1}{80} \theta^5
\!-\! \tfrac{1}{15} \pi^5 \; , 
\label{Ls5_series}
\\ && \hspace*{-7mm}
\LS{4}{1}{\theta}
= 
- 2 \sum_{n=1}^\infty \left[\frac{\cos (n \theta)}{n^3} + \theta \frac{\sin (n \theta)}{n^2}\right] S_1
- \tfrac{1}{48} \theta^2 (6 \pi^2 - 8 \pi\theta + 3 \theta^2)
+ \tfrac{1}{180} \pi^4 \;, 
\label{LS4_1_series}
\\ && \hspace*{-7mm}
\LS{5}{1}{\theta}
= 
\tfrac{3}{4} \sum_{n=1}^\infty \frac{\cos (n \theta)}{n^3} 
\Biggl(
\frac{6}{n^2} 
\!-\! \frac{8}{n} S_1
\!+\! 4 \left[ S_1^2 \!-\! S_2 \right]
\!-\! (\pi^2 \!-\! \theta^2) 
\Biggr)
\nonumber \\ && \hspace*{-2mm}
+ 3 \sum_{n=1}^\infty \frac{\sin (n \theta)}{n^2} 
\Biggl(
\tfrac{1}{4} \theta (\pi \!-\! \theta)^2 
\!-\! \tfrac{1}{2} \frac{(2 \pi \!-\! \theta) }{n^2} 
\!+\! (\pi \!-\! \theta) \frac{S_1}{n} 
\!+\! \theta \left[ S_1^2 \!-\! S_2 \right]
\Biggr)
+ \tfrac{9}{2} \zeta_2 \zeta_3 
\!-\! \tfrac{9}{2} \zeta_5
\; , 
\label{LS5_1_series}
\\ && \hspace*{-7mm}
\LS{5}{2}{\theta}
= 
 2 \sum_{n=1}^\infty \frac{\sin (n \theta)}{n^2} S_1 
\Biggl( \frac{2}{n^2} \!-\! \theta^2 \Biggr)
\!-\! 4 \theta \sum_{n=1}^\infty \frac{\cos (n \theta)}{n^3} S_1
\!-\! \frac{\theta^3}{120}  \left( 
6 \theta^2 
\!-\! 15 \pi \theta
\!+\! 10 \pi^2
\right)
\; , 
\label{LS5_2_series}
\end{eqnarray}
where we have introduced a short notation 
$S_a \equiv S_a(n-1)$, 
and we used the general expression for the Stirling number of the first kind 
in terms of the harmonic sums \cite{Stirling}
\begin{eqnarray}
&& 
\ST{n}{0} = \delta_{0n} \; , 
\quad 
\ST{n}{1} = (n-1)! \; , 
\quad 
\ST{n}{2} = (n-1)! S_1 \; , 
\nonumber \\ && 
\ST{n}{3} = \tfrac{1}{2} (n-1)!  \left[S_1^2 - S_2 \right] \; , 
\quad 
\ST{n}{4} = \tfrac{1}{3!} (n-1)! \left[S^3_1 - 3 S_1  S_2  +  2 S_3 \right] \; , 
\nonumber \\ && 
\ST{n}{5} = \tfrac{1}{4!} (n-1)! \left[ 
S_1^4 
+ 8 S_1 S_3
+ 3 S_2^2 
- 6 S_4
- 6 S_1^2 S_2
\right] \; ,
\nonumber \\ && 
\ST{n}{6} = \tfrac{1}{5!} (n-1)! \left[ 
S_1^5
\!-\! 30 S_1 S_4 
\!-\! 20 S_2 S_3 
\!+\! 20 S_1^2 S_3 
\!-\! 10 S_1^3 S_2 
\!+\! 15 S_1 S_2^2  
\!+\! 24 S_5 
\right] \; .
\label{Stirling}
\end{eqnarray}
Using the results of paper \cite{DK1}
(see Eqs.~(A.9), (A.10), (A.14) in \cite{DK1})
we get from (\ref{Ls3_series})-(\ref{LS5_1_series}) the following
relations:
\begin{eqnarray}
&& 
\label{zoo_begin}
\sum_{n=1}^\infty \frac{ \sin \left(\tfrac{\pi}{3}n \right)}{n^2} S_1 = \tfrac{1}{54} \pi \zeta_2 \; , 
\\ && 
\sum_{n=1}^\infty \frac{ \cos \left(\tfrac{\pi}{3}n \right)}{n^3} S_1 = - \tfrac{23}{216} \zeta_4 \; , 
\\ && 
\sum_{n=1}^\infty \frac{ \sin\left(\tfrac{\pi}{3}n \right)}{n^2} \Biggl[ S_1^2 - S_2 \Biggr]
=  
  2 \Cl{4}{\tfrac{\pi}{3}}
- \tfrac{2}{3} \zeta_2 \Cl{2}{\tfrac{\pi}{3}}
- \tfrac{2}{9} \pi \zeta_3 
\; , 
\\ && 
 3 \sum_{n=1}^\infty 
\frac{ \cos \left(\tfrac{\pi}{3} n \right)}{n^3} \Biggl[ S_1^2 \!-\! S_2  \!-\! 2 \frac{S_1}{n} \Biggr] 
\!+\!  2 \pi \sum_{n=1}^\infty \frac{ \sin \left(\tfrac{\pi}{3} n \right)}{n^3} S_1 
= 
2 \pi \Cl{4}{\tfrac{\pi}{3}}
\!-\! \tfrac{4}{3} \zeta_2 \zeta_3
\!-\! \tfrac{7}{3} \zeta_5
\; , 
\\ && 
\sum_{n=1}^\infty \frac{ \cos \left(\tfrac{\pi}{2} n \right)}{n^3} S_1 = 
- \tfrac{181}{256} \zeta_4 
+ \tfrac{5}{192} \ln^4 2
- \tfrac{5}{32} \zeta_2 \ln^2 2 
+ \tfrac{35}{64} \zeta_3 \ln 2
+ \tfrac{5}{8} \Li{4}{\tfrac{1}{2}} \;,  
\\ && 
3 \sum_{n=1}^\infty 
\frac{ \cos \left(\tfrac{\pi}{2} n \right)}{n^3} \Biggl[ S_1^2 \!-\! S_2  \!-\! 2 \frac{S_1}{n} \Biggr] 
+ \tfrac{3}{2} \pi 
\sum_{n=1}^\infty \frac{ \sin \left(\tfrac{\pi}{2} n \right)}{n^3} S_1 
= 
\tfrac{3}{2} \pi \Cl{4}{\tfrac{\pi}{2}}
-  \tfrac{93}{1024} \zeta_5 
- \tfrac{315}{256} \zeta_2 \zeta_3 
\nonumber \\ && \hspace{2mm}
- \tfrac{1}{16} \ln^5 2
+ \tfrac{5}{16} \zeta_2 \ln^3 2
- \tfrac{105}{128} \zeta_3 \ln^2 2
- \tfrac{15}{8} \Li{4}{\tfrac{1}{2}} \ln 2 
- \tfrac{15}{8} \Li{5}{\tfrac{1}{2}} \; . 
\label{zoo:2}
\end{eqnarray}


\end{document}